\documentclass[journal=jacsat,manuscript=article]{achemso}
\usepackage[]{achemso}
\setkeys{acs}{etalmode = truncate}
\setkeys{acs}{maxauthors = 10}

\usepackage{chemformula} 
\usepackage[T1]{fontenc} 

\usepackage{subfig}

\newcommand{\cmmo}{cm$^{-1}$}      
\newcommand{\hhcl}{H$_2$Cl$^+$ -- H$_2$}  
\newcommand{\ie}{{\it i.e.}}

\newcommand{\etal}{{\it et al.}}

\author{S\'{a}ndor Demes}
\email{demes.sandor@atomki.hun-ren.hu}
\affiliation[Univ Rennes]
{Univ Rennes, CNRS, IPR (Institut de Physique de Rennes) - UMR 6251, F-35000 Rennes, France}
\altaffiliation{Current address: HUN-REN Institute for Nuclear Research, Bem square 18/c, H-4026 Debrecen,
Hungary}
\author{Dariusz K\k{e}dziera}
\affiliation[Troun University]
{Nicolaus Copernicus University in Toru\'{n}, Faculty of Chemistry, Gagarina 7, 87-100 Toru\'{n}, Poland}
\author{Alexandre Faure}
\affiliation[Univ Grenoble-Alpes]
{Universit\'{e} Grenoble Alpes, CNRS, IPAG, F-38000 Grenoble, France}
\author{Fran\c{c}ois Lique}
\affiliation[Univ Rennes]
{Univ Rennes, CNRS, IPR (Institut de Physique de Rennes) - UMR 6251, F-35000 Rennes, France}
\email{francois.lique@univ-rennes.fr}

\title[H$_2$Cl$^+$ collision with H$_2$]
  {Rotational excitation cross sections for chloronium based on a new 5D interaction potential with molecular hydrogen \footnote{Accepted in {\it J.~Phys.~Chem.~A}  on 12/12/2024 (\url{https://doi.org/10.1021/acs.jpca.4c07467})}}

\abbreviations{CC,PES,CS}

\begin{document}

%

\begin{abstract}
    Chloronium (\ch{H2Cl+}) is an important intermediate of Cl-chemistry in space. The accurate knowledge of its collisional properties allows a better interpretation of the corresponding observations in interstellar clouds and therefore a better estimation of its abundance in these environments. While the ro-vibrational spectroscopy of \ch{H2Cl+} is well known, the studies of its collisional excitation are rather limited and these are available for the interaction with helium atoms only. We provide a new 5-dimensional rigid-rotor potential energy surface for the interaction of \ch{H2Cl+} with \ch{H2}, calculated from explicitly correlated coupled cluster {\it ab initio} theory, which was fitted then with a set of analytical functions, allowing to perform scattering calculations using accurate quantum theories. We analyze the collision-energy-dependence of the rotational state-to-state cross sections and the temperature dependence of the corresponding thermal rate coefficients, with a particular attention on the collisional propensity rules. When comparing our results for collisions with \ch{H2} with those obtained with He as a colliding partner, we found very significant differences with non-linear scaling trends, which proves again that He is not a suitable proxy for collisions between hydride molecules and molecular hydrogen, the most abundant gas particle in the interstellar medium.
\end{abstract}

\section{\label{sec:Intro} Introduction}

Only seven chlorine-bearing species have been detected so far in the interstellar medium (ISM) \cite{thorwirth2024}. Nevertheless chemical models suggest that, despite their relatively low abundance, such halogen elements can be important in various regions of the ISM \cite{acharyya2017}. This is primarily due to their unique thermochemical affinity to form hydrides with strong chemical bonds such as \ch{HCl} or \ch{H2Cl+}, whose abundance in some interstellar molecular clouds can be comparable to those of widespread species such as \ch{H2O} \cite{wallstrom2019}. Atomic chlorine has some special properties, most notably a low ionization potential, which supports the formation of ionized \ch{Cl+} in the atomic phase that can initiate an active chemistry through reaction with \ch{H2}. Some recent astronomical observations showed that one of the products of these reactions, chloronium cation \ch{H2Cl+} is an important participant in interstellar chlorine chemistry \cite{lis2010}, where the most dominant contributions have been connected to diatomic species earlier. It can then react with electrons to either form simple HCl or neutral chlorine. The first interstellar detection of \ch{H2Cl+} by \citet{lis2010} showed that its abundance is significantly larger than it was supposed in earlier chemical models. \citet{neufeld2015} also derived a fractional abundance about a factor of 5 higher than astrochemical models. This has paved the way for several astronomical surveys for \ch{H2Cl+}, both in dense and diffuse clouds of the ISM. Chloronium has been detected in various environments, first through the {\it Herschel Space Observatory} in the Milky Way \cite{lis2010, neufeld2012, neufeld2015} then in extragalactic sources \cite{muller2014} and most recently in front of quasar molecular absorbers \cite{legal2017,wallstrom2019}. However, despite of these detections, there are still open question in interstellar chlorine chemistry \cite{Ritchey2023}, which might be answered through a better interpretation of the observations. For this, a proper astrophysical modelling is required, which takes into account the non-local thermodynamic equilibrium (\ie~non-LTE) effects of the environment through accurate radiative transfer models including excitation by both radiation and collisions, the latter being presently missing in the literature.

Only a few attempts have been made to reveal the dynamical aspects of \ch{H2Cl+} in collisions. First, \citet{legal2017} studied the \ch{HCl+ + H2 -> H2Cl+ + H} reaction from a quasi-classical trajectory (QCT) model, which was found to be the main path of formation of chloronium. The work was devoted to quantify the {\it ortho}-to-{\it para} ratio of \ch{H2Cl+} under diffuse cloud conditions, but the authors do not report state-selective rate coefficients, which would be important for a proper astrophysical modelling. The first state-to-state collisional data for the rotational excitation of chloronium has been reported very recently by \citet{mehnen2024}. The authors calculated cross sections and rate coefficients for \ch{H2Cl+} in collision with He, which is the second most abundant colliding partner in the ISM, often used to mimic interaction with \ch{H2}. Cross sections have been calculated for kinetic energies up to $1000$~\cmmo, and thermal rate coefficients are obtained from 10 to 150~K for rotational states involving $j \leq 9$. It should be stressed, however, that there are no data in the literature for collisions with \ch{H2}, which is the most abundant collider in translucent and dense clouds of the ISM, for example in the Orion Molecular Cloud 1 studied by \citet{neufeld2012}. There is also a lack of data for \ch{H2Cl+} in diffuse clouds, where collisions with atomic hydrogen and electrons dominate.

The vibrational spectra and the equilibrium structure of chloronium have been actively studied already in the 80s \cite{Lee1988,Botschwina1988}. Following the early works at the end of the past century, the most accurate spectroscopic properties of \ch{H2Cl+} have been identified through sub-mm measurements by \citet{araki2001} and also through calculations by \citet{afansounoudji2023}, where the authors found the following anharmonic vibrational frequencies: $\nu_1 = 2648.2$~\cmmo, $\nu_2 = 1184.6$~\cmmo~ and $\nu_3 = 2634.9$~\cmmo. It is important to note that these studies describe a series of isotopologues apart from the main \ch{H2^{35}Cl+} one, including \ch{H2^{37}Cl+} and \ch{HD^{35}Cl+}, while the latest study \cite{afansounoudji2023} also reports data on \ch{D2^{35}Cl+}, \ch{D2^{37}Cl+} and \ch{HD^{37}Cl+} species. Due to the heavy mass of the chlorine atom, there is no significant difference between the structure and rotational constants of \ch{^{35}Cl+}- and \ch{^{37}Cl+}-containing species, however the deuterated substitutes possess very different spectroscopic properties.

The aim of this work is to provide the first rotationally inelastic cross sections and thermal rate coefficients for the collision of \ch{H2Cl+} with \ch{H2} based on a new, accurate 5-dimensional (5D) potential energy surface (PES). The numerically exact close coupling scattering method has been applied for the collision dynamics calculations. The results are compared with the available data for the \ch{H2Cl+ + He} collision \cite{mehnen2024}. The paper is organized as follows. In Section~2\ref{sec:Meth}, the {\it ab initio} calculations and the potential energy surface details are discussed, Section~3\ref{sec:ScattMeth} describes the scattering calculation methodologies. In Section~4\ref{sec:ResDis}, we report the collisional cross sections and discuss their propensity rules, while our conclusions are drawn in Section~5\ref{sec:Concl}.

\section{\label{sec:Meth} Methods}

\subsection{\label{sec:Abinit} Ab initio calculations}

\begin{figure}
\includegraphics[width=0.6\linewidth]{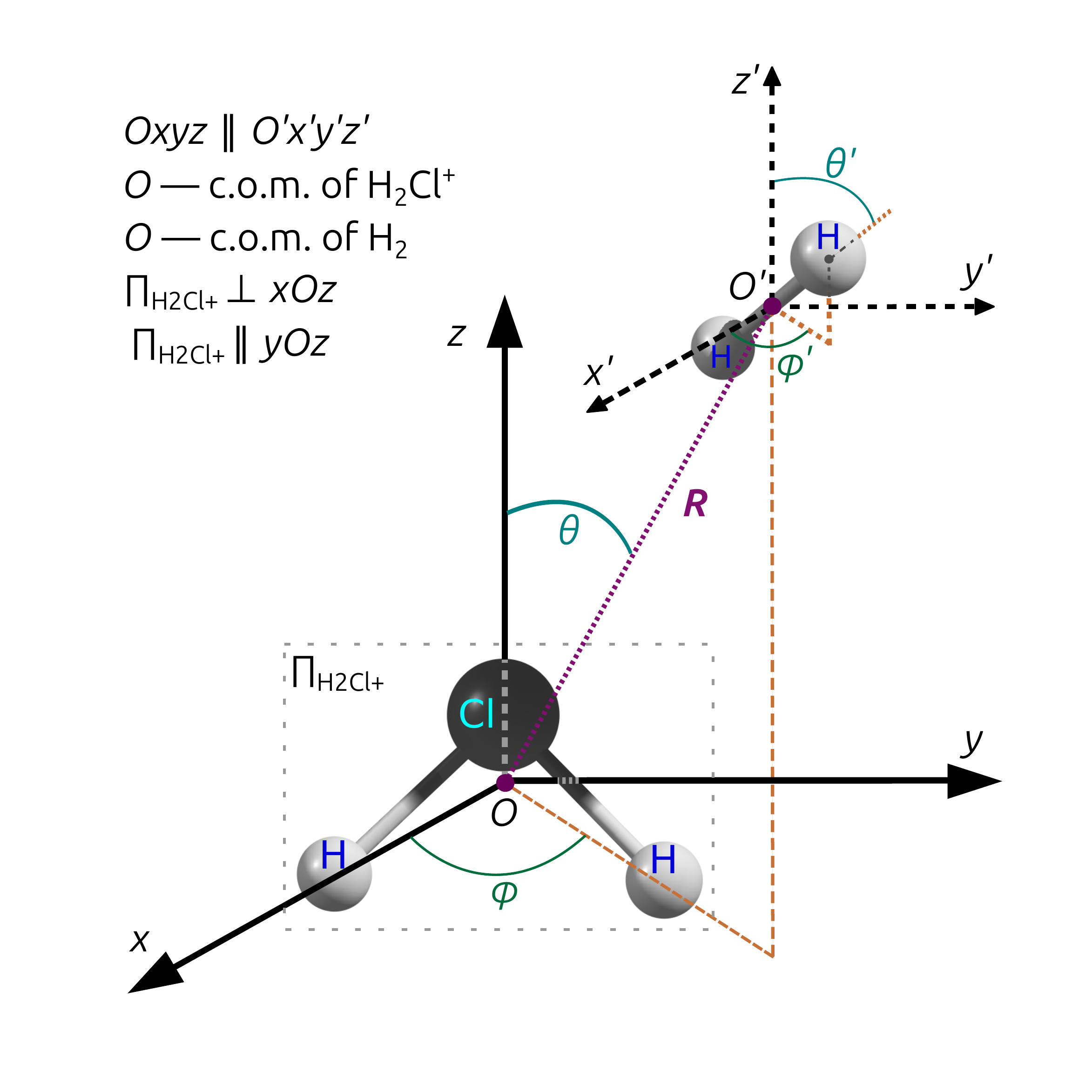}
\caption{The 5D coordinate system for the \hhcl collision.}
\label{fig:coordsys}
\end{figure}

For studying rotational excitation in low-energy molecular collision, first the ground-state electronic PES has to be determined. To describe the dynamics of the \hhcl~collision, a 5D Jacobi coordinate system has been used (Fig.~\ref{fig:coordsys}). The rigid rotor approximation have been used, which neglects the intramolecular vibrational motion, \ie~the Cl--H and H--H bonds were kept fixed. The PES is described then in terms of $R, \theta, \phi , \theta'$ and $\phi'$ coordinates. The center of the coordinate system ($O$ in Fig.~\ref{fig:coordsys}) is in the center of mass (c.o.m.) of \ch{H2Cl+}, while the $z$-direction points towards chlorine atom, and the \ch{H2Cl+} target molecule lies in the $yz$-plane. The $R$ radial parameter defines the intermolecular distance between the c.o.m.(s) of \ch{H2Cl+} and \ch{H2}. The $\theta$ and $\phi$ angles reflect to the position of \ch{H2} with respect to centre $O$, while $\theta'$ and $\phi'$ define the relative orientation of \ch{H2} around its own c.o.m.~($O'$). Similar notations and coordinate system were used earlier to study the rotational excitation in \ch{H2O - H2} \cite{Phillips1994,valiron2008,Faure_2007}, \ch{NH2 - H2} \cite{Bouhafs_2017} and \ch{H2S - H2} \cite{dagdigian2020} collisions. An $r_\mathrm{H-H} = 1.44874$ bohr bond length was considered for \ch{H2} (its ground-state equilibrium distance accurately calculated without using the Born–Oppenheimer approximation \cite{bubin2003}). For the \ch{H2Cl+} cation, the equilibrium geometry have been located following a full geometry optimization using the explicitly correlated coupled cluster theory with single and double excitation and triple corrections (CCSD(T)-F12) \cite{adler2007} using the \texttt{MOLPRO} quantum chemistry software package\cite{MOLPRO_brief,werner2012}. In all CCSD(T)-F12 calculations, the recommended augmented, correlation-consistent valence triple-$\zeta$ (aug-cc-pVTZ) basis set has been used with standard density fitting (DF) and resolution of the identity (RI) basis terms.\cite{dunning1989,weigend2002}. The optimized bond parameters of \ch{H2Cl+}, which has been used to construct the coordinate system, are $r_\mathrm{Cl-H} = 2.4688$ bohr and $\alpha_\mathrm{H-Cl-H} = 94.4^{\circ}$. These data agrees better than $2\%$ with respect to the zero-point averaged structure of \ch{H2Cl+} obtained from precise measurements ($r = 2.4955$ bohr; $\alpha = 94.24^{\circ}$) \cite{araki2001} and calculations ($r = 2.504$ bohr; $\alpha = 94.33^{\circ}$) \cite{afansounoudji2023}. Throughout the paper, atomic units (a.u.) are used for distances (1~a.u. (bohr) $\approx 0.529177$ \AA), and wavenumbers (cm$^{-1}$) for energies (1~cm$^{-1} \approx 1/219474.624$ hartree).

We calculated the \textit{ab initio} energies for a total of 117 000 single-point geometries, where a set of 3000 randomly generated angular configurations for $\{ \theta, \phi , \theta', \phi' \}$ has been applied for 39 radial distances, spanning the space between $R=3.25$ and $R=40.0$~a.u. The radial step size $\Delta R$ has been varied as follows: $\Delta R=0.25$~a.u. below $8$~a.u., $\Delta R=0.5$~a.u. between $8$ and $11$~a.u., $\Delta R=1.0$~a.u. between $11$ and $20$~a.u., and we considered a few long-range points at $R=22, 25, 32.5, 40$~a.u. The CCSD(T)-F12b/aug-cc-pVTZ theory has been used for all electronic structure calculations, which is well-suited for the accurate description of ground-state interaction potentials, as it was shown, for example, in our previous works \cite{Demes_2020,Demes_2024}. The calculated energies have been corrected for basis set superposition error using the counterpoise procedure \cite{Boys1970}. To eliminate the size-inconsistency of the CCSD(T)-F12 method, the single-point energies calculated at $R=100$~a.u. have been subtracted from the particular interaction energies that match with the same angular configuration.

\subsection{\label{sec:PESfit} Analytical fit of the potential energy surface}
In order to efficiently use the PES for time-independent quantum close-coupling scattering calculations, it is convenient to employ a bispherical harmonic expansion. We used the convention proposed for the similar \ch{H2O - H2} collision \cite{Phillips1994,valiron2008}, where

\begin{equation}
V(R, \theta, \phi , \theta',\phi') = \sum_{l_1m_1l_2l} v_{l_1m_1l_2l} (R) \overline{t}_{l_1m_1l_2l} (\theta, \phi , \theta',\phi')
\label{eqn:main},
\end{equation}
where the normalized spherical tensor $\overline{t}_{l_1m_1l_2l} (\theta, \phi , \theta',\phi')$ is defined as
\begin{eqnarray}
\notag
\overline{t}_{l_1m_1l_2l} (\theta, \phi , \theta',\phi') && = \alpha_{l_1m_1l_2l} \big(1+\delta_{m_10} \big)^{-1} \sum_{r_1r_2}
    \begin{pmatrix}
    l_1 & l_2 & l \\
    r_1 & r_2 & r
    \end{pmatrix} \times Y_{l_2r_2}(\theta',\phi') Y_{lr}(\theta,\phi) \\
\notag
&& \times \big[ \delta_{m_1r_1} + (-1)^{l_1+m_1+l_2+l} \delta_{-m_1r_1} \big]
\label{eqn:tensor},
\end{eqnarray}
with a normalization factor \( \alpha_{l_1m_1l_2l} = \big[2(1+\delta_{m_10})^{-1} (2l_1 +1)^{-1} \big] ^{-1/2} \) (see Valiron \etal \cite{valiron2008} for more details). The $l_1$, $l_2$ and $l$ indices here refer to the tensor ranks of the angular dependence of the \ch{H2Cl+} orientation, the \ch{H2} orientation, and the collision vector orientation, respectively. The symmetry with respect to reflection of \ch{H2} in the \ch{H2Cl+} plane is taken into account through the phased sum over $\pm m_1$, with $|m_1| \leq l_1$, which is further restricted to be even due to the $C_{2v}$ symmetry of \ch{H2Cl+}. 
The homonuclear symmetry of \ch{H2} similarly constrains $l_2$ for even integers only. There are several benefits of using such expansion. First, the interaction potential is invariant to the inversion of the coordinates around the origin, and second, the quantization axis for the PES expansion is the same as that for the rotational Hamiltonian in the molecular-frame coordinate system (Fig.~\ref{fig:coordsys}) \cite{dagdigian2020}.

At each intermolecular separation $R$, the PES was developed over the 3000 angular orientations according to Eq.~(\ref{eqn:main}), and was fit to the corresponding {\it ab initio} energies to find the $v_{l_1m_1l_2l} (R)$ expansion coefficients, using a linear least-squares procedure. For a reasonably accurate fit, we selected anisotropies up to $l_1 = 16$ and $l_2 = 6$, resulting in a total of 1998 angular functions. The significant terms were selected then iteratively, using a Monte Carlo error estimator described in details by~\citet{rist2012}, resulting in a final set of 228 expansion functions.
The root mean square (RMS) error of the fit is better than 1~cm$^{-1}$ in the long-range ($R=40.0$~bohr) and minimum (mostly attractive) region of the PES, \ie~at intermolecular distances $R>5.0$, where the mean error on the expansion coefficients $v_{l_1l_2m_1m_2}(R)$ was also found to be $<1$~cm$^{-1}$. We then employed a standard cubic spline radial interpolation of the fitted expansion coefficients over the whole intermolecular range, which was smoothly connected to physical extrapolations both in the short and long ranges, as proposed by Valiron \etal \cite{valiron2008}. We used a standard exponential extrapolation technique in the short range $v_{l_1l_2m_1m_2}(R) = A \exp(-\alpha R)$, with a smooth transition domain between $3.25$~bohr and $4.0$~bohr. In the long range, the most significant coefficients have been extrapolated using the usual inverse power law formula $v_{l_1l_2m_1m_2}(R) = B/R^{\beta}$, with a smooth transition domain from $32.5$ to $40.0$~bohr. The $A,B$ factors and $\alpha, \beta$ exponents are determined by interpolation to the {\it ab initio} data corresponding the three closest distances to the short- and long-ranges, respectively. The switch function in these domains is defined the same way as \citet{valiron2008} (see Eqn.~(10) therein):
\begin{equation}
f(x) = \frac{1}{2} \biggl \{ 1- \cos \biggl[ \frac{1-\cos(x\pi)}{2} \pi \biggl] \biggr \}
\label{eqn:switch}.
\end{equation}
With this procedure a fully analytical PES has been obtained, which was implemented then in the \texttt{MOLSCAT} molecular scattering code \cite{Hutson_2019}. Based on systematic dynamical calculations, we further reduced the number of terms, selecting only the most significant $142$ angular expansion terms, which ensures an accuracy better than $1\%$ with respect to the original (228-terms) basis. This truncation could save a reasonable amount of CPU time with a negligible loss of numerical precision on the cross sections.

\begin{figure}
\centering
    \subfloat[\label{subfig_a:ContThR}]{{\includegraphics[width=0.49\textwidth]{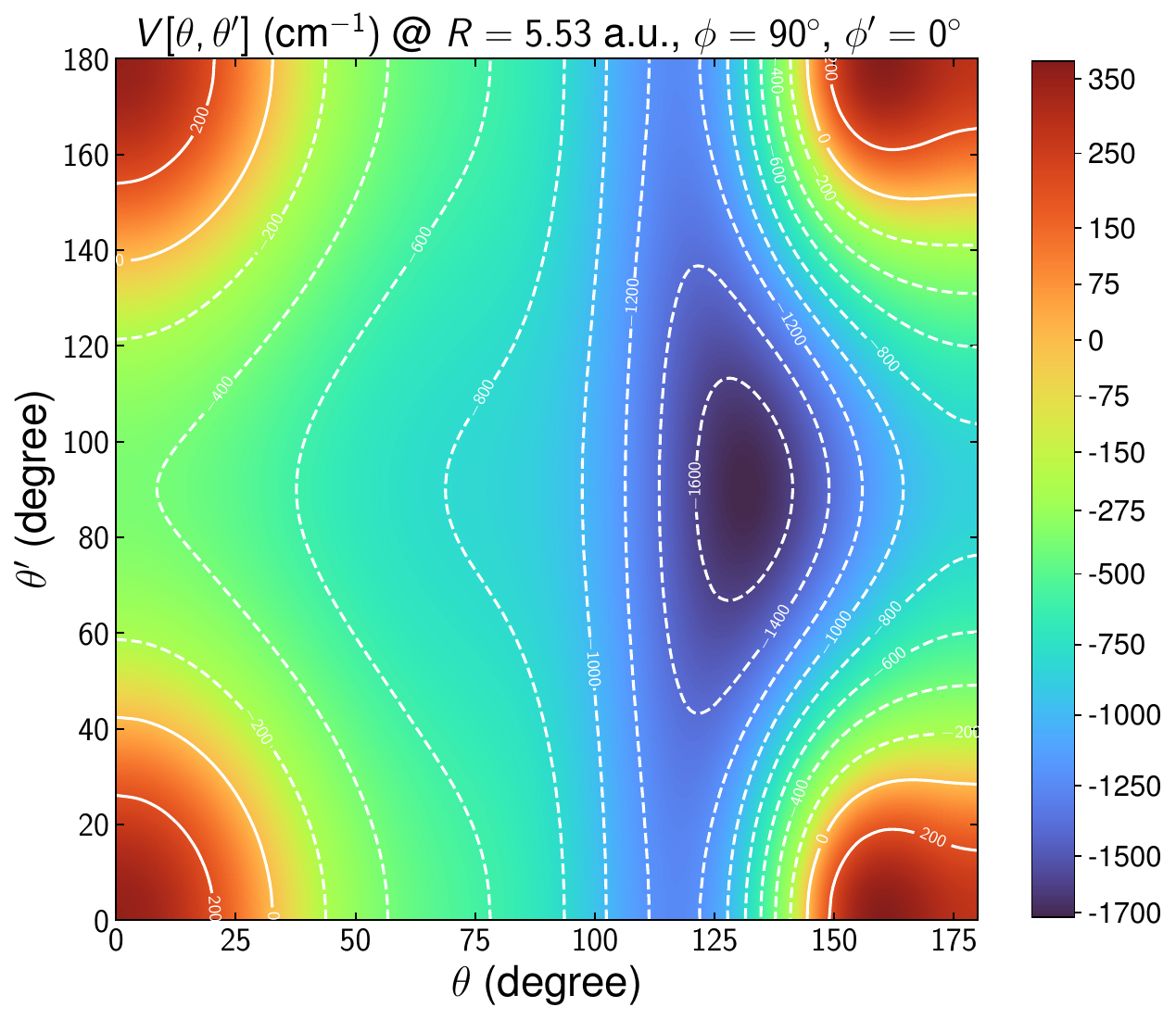} }}%
    \subfloat[\label{subfig_b:ContPhR}]{{\includegraphics[width=0.49\textwidth]{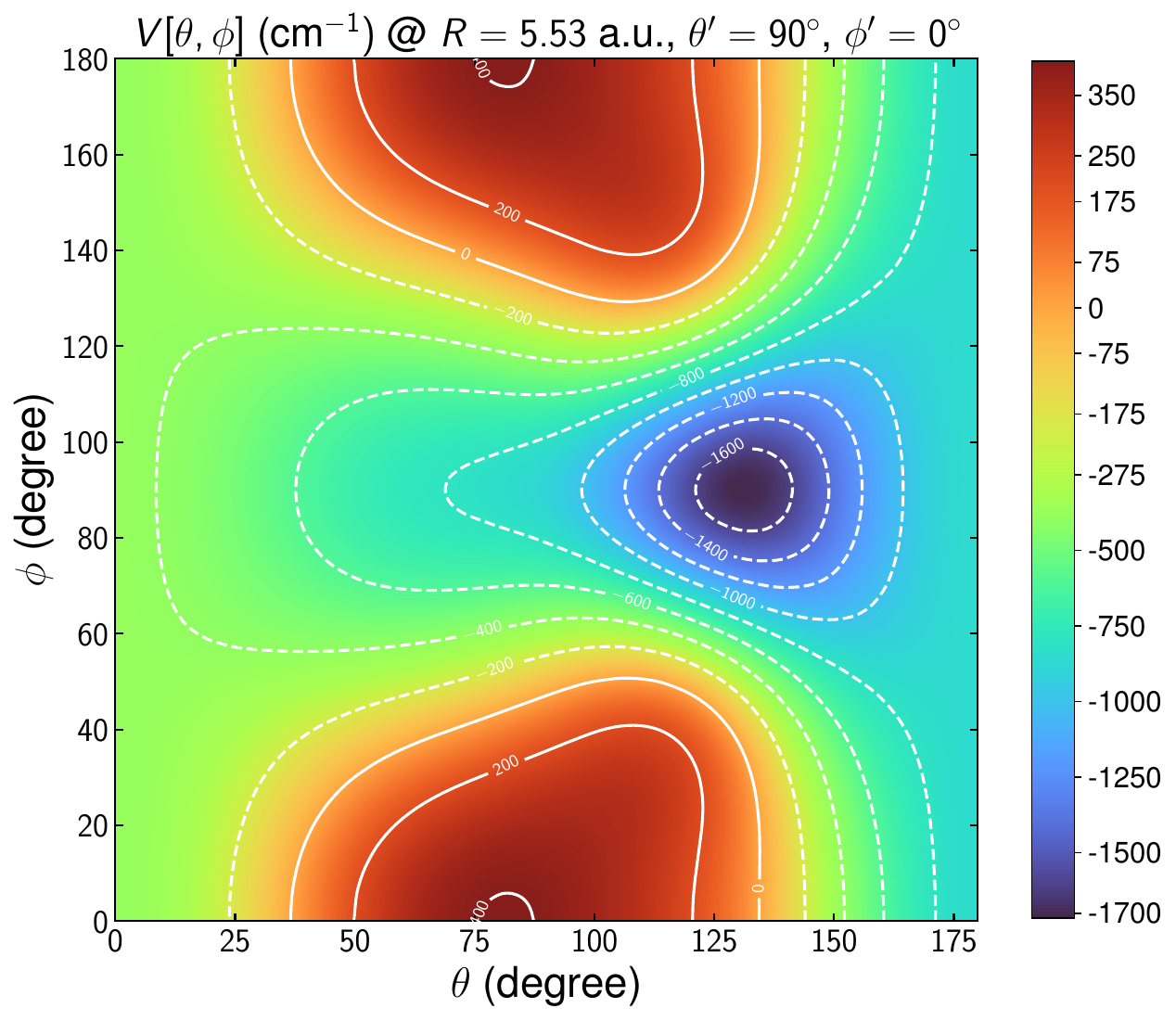} }}%
\caption{Angular dependence contour plots for the \hhcl~interaction potential at fixed radial distance and angular positions.}
\label{fig:contourPES1ab}
\end{figure}

We also examined the behaviour of the analytical PES visually by constructing contour plots while fixing different dimensions. In all cases, the contour plots were determined using the full set of 228 angular coefficients. The global minimum (GM) of the PES is located at $R \simeq 5.53$~a.u., $\theta \simeq 131.22^{\circ}$, $\phi = 90^{\circ}$, $\theta' = 90^{\circ}$, and $\phi' = 0^{\circ}$, with a deep well of about $-1718$~\cmmo. This configuration corresponds to an \ch{H2} orientation that is perpendicular to the plane of the \ch{H2Cl+} target (plane $\Pi_\mathrm{H2Cl+}$ in Fig.~\ref{fig:coordsys}), while being also perpendicular to the intermolecular vector $R$. The resulting minimum-energy geometry refers to a typical van der Waals-type complex that can be associated to the quadrupole-quadrupole interaction between \ch{H2Cl+} and \ch{H2}. The GM position and depth was also verified through CCSD(T)-F12b/aug-cc-pVTZ calculations, which is in a perfect agreement ($<0.01\%$) with the analytical PES. The contours of the interaction potential in the $\theta$  {\it vs.} $\theta'$ as well as in $\theta$  {\it vs.} $\phi$ coordinates are shown in Figs.~\ref{fig:contourPES1ab}(a) and (b), respectively, at \hhcl~intermolecular separation and fixed angles that correspond to the position of the GM. These plots help in understanding some aspects of the angular anisotropy of the PES. The interaction is dominantly attractive here, as seen in Fig.~\ref{fig:contourPES1ab}(a), apart from a rather narrow region close to boundary angles ($\theta, \theta' \simeq 0^{\circ} \pm 30^{\circ}$). The minimum region around $\theta \simeq 130^{\circ}$ and $\theta'=90^{\circ}$ is not too sharp, the attractive potential in this region is rather invariant to \ch{H2} rotation in the $\theta'$-dimension. It is even more visible as leaving the well, for example the $V=1200$~\cmmo~contour line is quasi-independent on $\theta'$. There is a perfect left-right symmetry with respect to $\theta'=90^{\circ}$, which is due to the 2-fold symmetry of the hydrogen molecule (note that $\theta'$ defines the orientation of \ch{H2} relative to the $z$-axis). The angular anisotropy is significant, but not extremely large ($V_\mathrm{max} \approx 370$ \cmmo) in these polar angle dimensions.

Fig.~\ref{fig:contourPES1ab}(b) is suitable to examine the angular dependence of the PES on the position of \ch{H2}. The interaction is dominantly attractive apart from a wider region centred at $\phi=0^{\circ}$ and around $\theta \simeq 80^{\circ}$. The angular anisotropy is found to be strong with respect to rotation along $\phi$. For example, when fixing $\theta \sim 120^{\circ}$, spanning through both attractive and repulsive regions of the PES takes place during a $\phi=\pi/2$ rotation. Due to this, the well region is more confined in these coordinates. The interaction potential is invariant to the azimuthal angle in the vicinity of $\theta = 0^{\circ}$ and $\theta = 180^{\circ}$ polar angles, as obvious from the definition of the coordinate system (the c.o.m. of \ch{H2} is on the $z$-axis). Due to the $C_{2v}$ symmetry of the chloronium, a perfect symmetry is found with respect to $\phi=90^{\circ}$. It is important to highlight here, that the analytical fit was carried out based on {\it ab initio} points on a randomly generated angular grid, which covers the full range of angles with $\phi, \phi' = [0^{\circ},360^{\circ}]$ and $\theta,\theta' = [0^{\circ},180^{\circ}]$. Therefore, the observed bilateral symmetries together with the smooth contour lines reflect on the good quality of the angular expansion.

\begin{figure}
\includegraphics[width=0.70\linewidth]{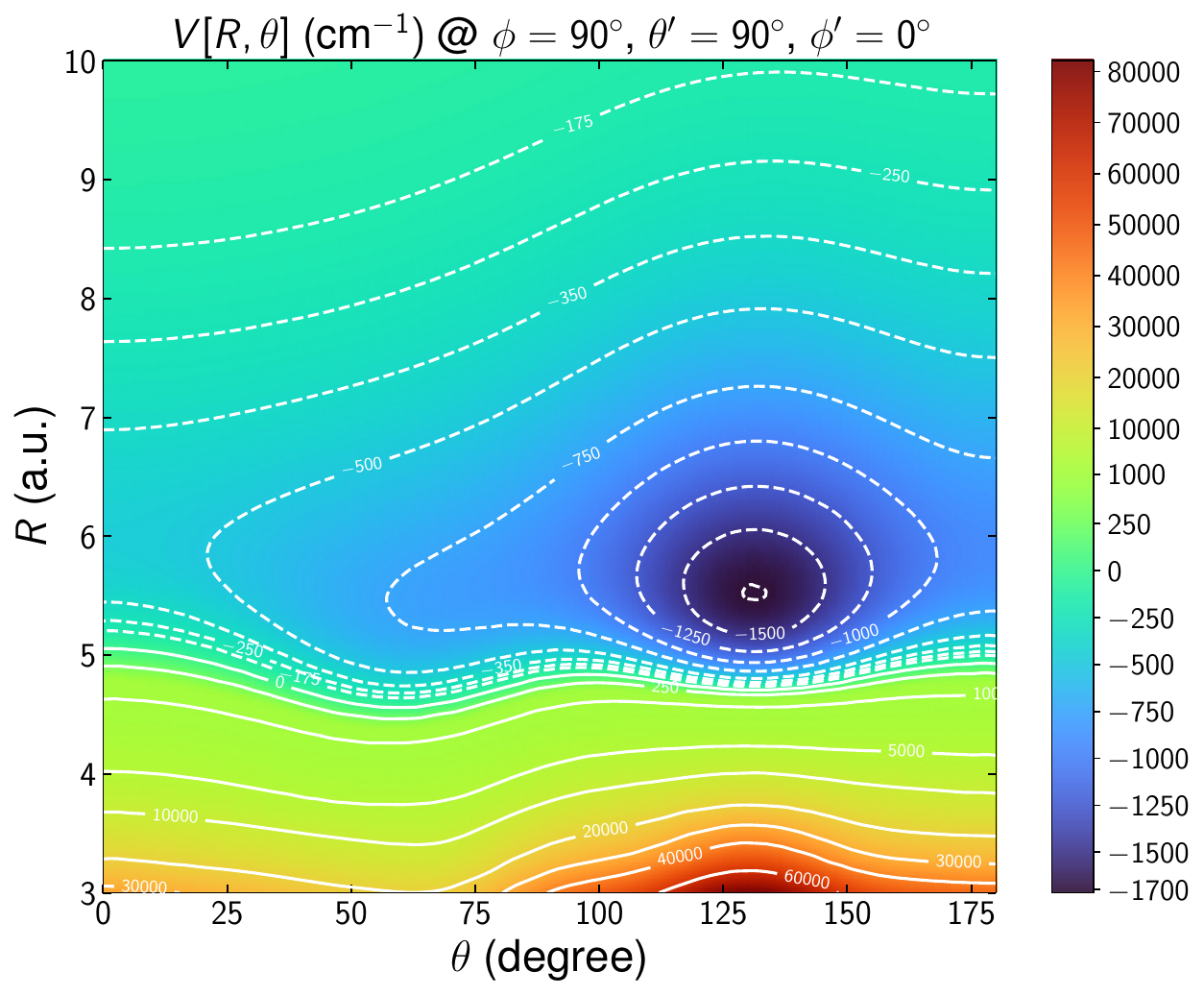}
\caption{Radial dependence contour plots for the \hhcl~interaction potential at fixed angular positions, which correspond the position of the global minimum.}
\label{fig:contourPES2}
\end{figure}

We also examined the radial dependence of the \hhcl~interaction potential. Fig.~\ref{fig:contourPES2} displays the $\theta$  {\it vs.} $R$ contour plot of the PES at fixed $\phi = 90^{\circ}$, $\theta' = 90^{\circ}$, and $\phi' = 0^{\circ}$ angles, which correspond to the GM orientation. The global well is very pronounced, located at $R \simeq 5.53$~$a_0$ and $\theta \simeq 131.22^{\circ}$. The interaction potential exhibits very large gradients of energy in the radial dimension. There are no signs of additional secondary minima and the overall $R$-dependence of the PES is rather symmetric (at fixed $\theta$). It is worth noting the lack of barriers towards the well, which indicates an efficient formation of the collisional complex.


\subsection{\label{sec:ScattMeth} Scattering calculations}

The analytically fitted PES was implemented in the \texttt{MOLSCAT} quantum scattering code \cite{Hutson_2019}. We calculated then the state-to-state cross sections for the rotational excitation of \ch{H2Cl+} due to collision with \ch{H2} from the numerically exact time-independent close-coupling (CC) quantum scattering method \cite{Arthurs_Dalgarno1960}. Similar calculations has been carried out for the \ch{H2O - H2} \cite{valiron2008} and \ch{H2S - H2} \cite{dagdigian2020} ``asymmetric top + diatom'' collision systems. The rotational excitation of \ch{H2Cl+} has been also studied recently in collision with He atoms, based on a 3D PES \cite{mehnen2024}.

Both colliders has two nuclear spin isomers: {\it ortho} ($o$) and {\it para} ($p$). \ch{H2Cl+} is an asymmetric top species, where the rotational levels are defined as $j_{k_a,k_c}$, where $j$ is the main rotational quantum number and $k_a,k_c$ are the projection of $j$ on the $A$ and $C$ rotational axes (note that according to spectroscopic notation, axis $A$ corresponds to the smallest moment of inertia of the molecule, while $C$ is the largest). The sum of the projection quantum numbers $k_a + k_c$ defines the nuclear spin configuration of chloronium, so that the odd sum refers to {\it ortho} and the even sum define the {\it para} states. In the case of \ch{H2}, the even rotational quantum numbers $(j_2 = 0,2,\dots)$ refer to {\it para}-configuration, and odd ones $(j_2 = 1,3, \dots)$ refer to {\it ortho}-states. We studied the collision of both {\it ortho-} and {\it para-}\ch{H2Cl+} with ground-state {\it para-}\ch{H2} $(j_2 = 0)$ (which is the dominant form of molecular hydrogen at $T<100$~K). In the case of chloronium, all transitions between levels with internal energies below $125$ \cmmo~ have been calculated, which covers all states with $j\leq3$ and some of those with $j=4$ (note that the lowest vibrational frequency of \ch{H2Cl+} is $\nu_2 = 1184.6$~\cmmo \cite{afansounoudji2023}, so the rigid rotor approximation is suitable for the current study). The rotational structure of \ch{H2Cl+} has been studied using a sub-mm spectroscopy technique,\cite{araki2001} where the following rotational constants have been found: $A = 337.3519$~GHz, $B = 273.5870$~GHz and $C = 148.1004$~MHz. We applied these spectroscopic parameters in the scattering calculations, and the calculated transition frequencies were in a below-percent agreement with those provided by \citet{araki2001} The full list of \ch{H2Cl+} rotational levels that have been involved in the current work are given in Table~\ref{tab:rot-levels}. The reduced mass of the [\hhcl] complex is  1.9115~amu.

\begin{table*}
	\centering
    \caption{List of the rotational levels for {\it o/p-}\ch{H2Cl+}, which are covered by the present work. The energies are taken from the sub-mm spectroscopy measurement by \citet{araki2001}}
    \label{tab:rot-levels}
    \begin{tabular}{rrr|rrr}
        \hline
        \multicolumn{3}{c}{\textbf{\textit{para-}\ch{H2Cl+}}} &  \multicolumn{3}{c}{\textbf{\textit{ortho-}\ch{H2Cl+}}}\\
        \hline
        state  & rotational  & rotational & state  & rotational  & rotational \\
        label & state $j_{k_a,k_c}$  & energy [\cmmo] & label & state $j_{k_a,k_c}$ & energy [\cmmo] \\
        \hline
         1.  &  $0_{0,0}$	 &     0.0000   &   1. &  $1_{0,1}$  &   14.0630  \\
         2.  &  $1_{1,1}$	 &    16.1916   &   2. &  $1_{1,0}$  &   20.3753  \\
         3.  &  $2_{0,2}$	 &    39.4931   &   3. &  $2_{1,2}$  &   40.1351  \\
         4.  &  $2_{1,1}$	 &    52.6719   &   4. &  $2_{2,1}$  &   59.0406  \\
         5.  &  $2_{2,0}$	 &    61.7215   &   5. &  $3_{0,3}$  &   74.5222  \\
         6.  &  $3_{1,3}$	 &    74.6554   &   6. &  $3_{1,2}$  &   98.3224  \\
         7.  &  $3_{2,2}$	 &   101.2237   &   7. &  $3_{2,1}$  &  111.0014  \\
         8.  &  $4_{0,4}$	 &   119.2121   &   8. &  $4_{1,4}$  &  119.2360  \\
         9.  &  $3_{3,1}$	 &   123.5558   &   9. &  $3_{3,0}$  &  124.9189  \\
        \hline
    \end{tabular}
\end{table*}

The full CC calculations can be optimised by truncating some of the variables, in particular the rotational basis size ($j_\mathrm{max}$) and largest total angular momentum ($J_\mathrm{tot}$). For this, we performed systematic convergence tests at various collision energies. We aimed at a good precision, so the threshold criteria to find convergence was set to a $0.1\%$ mean error for $j_\mathrm{max}$ and $0.005\%$ for $J_\mathrm{tot}$. The highest rotational levels which were included in the scattering calculations are those with $j_\mathrm{max} = 15$ and the largest angular momenta reached $J_\mathrm{tot} = 90$. Further optimisation can be achieved by restricting the numerical propagation space, in particular the initial ($R_\mathrm{min}$) and final radial distances ($R_\mathrm{max}$) and the switching point between the accurate log-derivative and the Airy propagators ($R_\mathrm{mid}$), for which we used identical criteria ($<0.1 \%$) to reach convergence. To properly resolve the resonance behaviour of the cross sections, the energy step size has been chosen as small as $0.1$~\cmmo~at all considered collision energies, from $0.1$~\cmmo~up to $500$~\cmmo. It is worth to mention that calculations involving higher \ch{H2} levels $(j_2=1,2,...)$ are computationally not feasible due to the large (142-term) expansion of the PES, which results in an extremely large number of coupled channels and makes CC-calculations demanding even when restricted to collision with spherical $p$-\ch{H2}~$(j_2=0)$, as used here. Some non-negligible uncertainties could therefore arise from the exclusion of the $j_2>0$ levels from the calculations. However, the estimated error due to the truncation of the \ch{H2} rotational basis should be less than $10\%$, and usually a few percent at most. On the other hand, as has been shown in the case of collisional excitation of various interstellar cations (see for example our previous work \citet{Demes_2022}), the interaction with $o$-\ch{H2}~$(j_2=1)$ or excited $p$-\ch{H2}~$(j_2=2)$ leads to very similar cross sections to those with ground-state $p$-\ch{H2}, so that calculations explicitly targeting these levels are not relevant for use in astrophysical models.

The calculated cross sections allowed to obtain thermal rate coefficients up to $50$ K (mostly $p$-\ch{H2} dominated) temperatures by integrating over a Maxwell-Boltzmann distribution of relative velocities:

\begin{equation}
    k_{\mathrm{i} \rightarrow \mathrm{f}}(T) = \left(\frac{8}{\pi\mu k_\mathrm{B}^3 T^3}\right)^\frac{1}{2} \int_{0}^{\infty} \sigma_{\mathrm{i} \rightarrow \mathrm{f}}(E_\mathrm{c})  E_\mathrm{c} \exp \left( {-\frac{E_\mathrm{c}}{k_\mathrm{B}T}} \right) dE_\mathrm{c} ,
    \label{eq:rates}
\end{equation}
where $E_\mathrm{c}$ is the collision or kinetic energy (note that the sum of kinetic and rotational energies defines the total energy), $\sigma_{\mathrm{i} \rightarrow \mathrm{f}}$ is the cross section for transition from initial state (i) to final state (f), $\mu$ is the reduced mass of the complex and $k_\mathrm{B}$ is the Boltzmann constant.

According to our estimations, the accuracy of the cross sections provided in this work should be below a few percents in general, and is always better than $\sim 20\%$. The most significant errors are due to the uncertainties from the {\it ab initio} calculations, from the analytical fit at short-range distances and due to the restricted rotational basis of $p$-\ch{H2}~$(j_2=0)$.

\section{\label{sec:ResDis} Results \& Discussion}

\begin{figure}
\centering
    \subfloat[\label{subfig_a:CS_dj1o}]{{\includegraphics[width=0.49\textwidth]{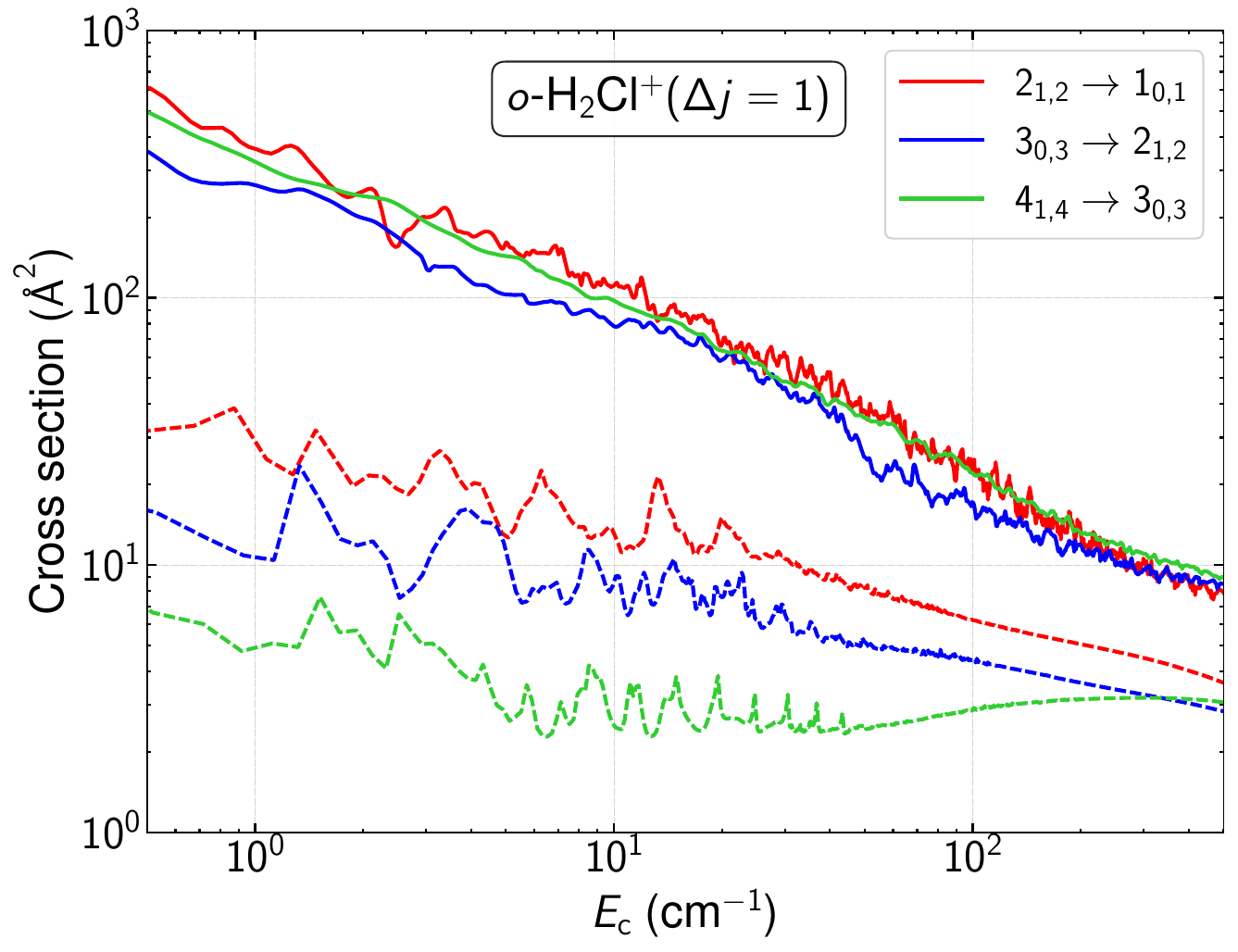} }}%
    \subfloat[\label{subfig_b:CS_dj1p}]{{\includegraphics[width=0.49\textwidth]{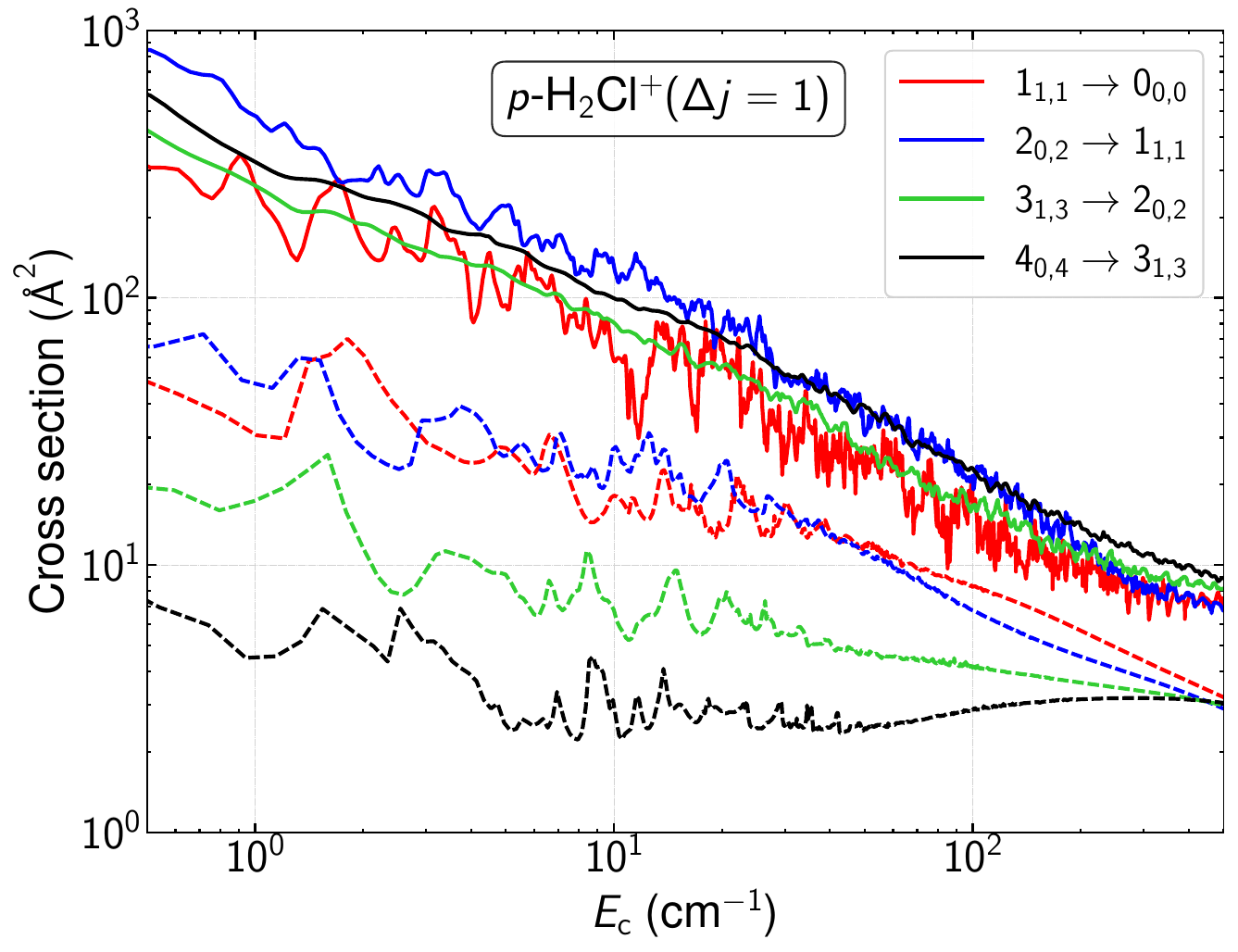} }}%
\caption{Collision-energy-dependence of the cross sections for $\Delta j = 1$ rotational de-excitation transitions of {\it ortho-}\ch{H2Cl+} (a) and {\it para-}\ch{H2Cl+} (b) due to collision with ground-state {\it para-}\ch{H2} (solid lines, current results) and He projectiles (dash-dot lines, reproduced from \citet{mehnen2024}).}
\label{fig:CS_dj1}
\end{figure}

Fig.~\ref{fig:CS_dj1}.(a)-(b) shows the collision-energy-dependence of the cross sections for {\it ortho}- and {\it para}-\ch{H2Cl+} (respectively) for some of those rotational de-excitation processes, where the main rotational quantum number decreases by one ($\Delta j = 1$). Our cross sections calculated for collision with $p$-\ch{H2} are compared with the corresponding data for the \ch{H2Cl+ + He} collision \cite{mehnen2024}. It is worth to mention again that in the scattering calculations only the rotational ground-state of \ch{H2} has been considered $(j_2=0)$, which has a spherically symmetric nature similarly to He. As can be seen in the plots, very significant differences are observed between all cross sections which correspond to the same transitions at all collision energies $E_\mathrm{c}$. Our cross sections with \ch{H2} are typically an order of magnitude larger than those calculated by \citet{mehnen2024} with helium colliding partner. While the general collision energy dependence is similar, we found some other significant discrepancies: the magnitudes of the cross sections in the case of \ch{H2} are very close to each other, while in the case of the He projectile significant differences (factors of about $2-5$) are observed at lower energies, which tends to disappear as the collision energy increases. The general trends are similar both in the case of $o$- and $p$-\ch{H2Cl+}, which is due to the deep well of the PES that favours coupling between levels with larger energy differences. It is worth to highlight also that the cross sections with \ch{H2} involve a dense structure of Feshbach- and shape resonances, especially the fundamental $1_{1,1} \rightarrow 0_{0,0}$ transition, which is also connected to large well depth ($\sim 1718$~\cmmo). For comparison, the global minimum of the \ch{H2Cl+ + He} PES is only about $260$~\cmmo~and there are no resonances observed for this system above $\sim 100$~\cmmo~collision energies \cite{mehnen2024}. Overall, the $\Delta j = 1$ cross sections for \ch{H2Cl+} are large with a relatively strong collision-energy-dependence, in particular they decrease by factors of $\sim 50-80$ as $E_\mathrm{c}$ is increasing from $0.5$ to $500$~\cmmo.

\begin{figure}
\centering
    \subfloat[\label{subfig_a:CS_pro}]{{\includegraphics[width=0.49\textwidth]{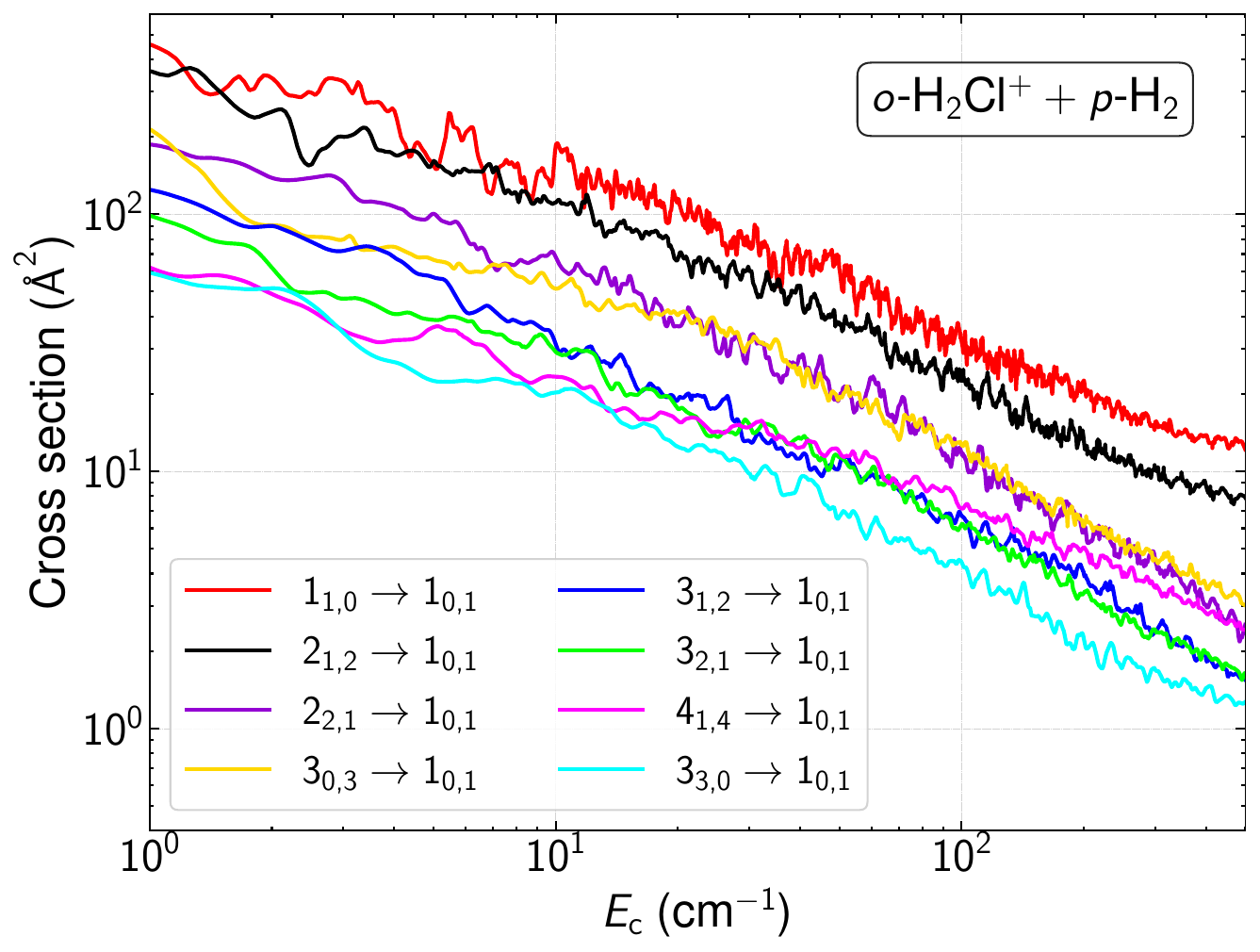} }}%
    \subfloat[\label{subfig_b:CS_prp}]{{\includegraphics[width=0.49\textwidth]{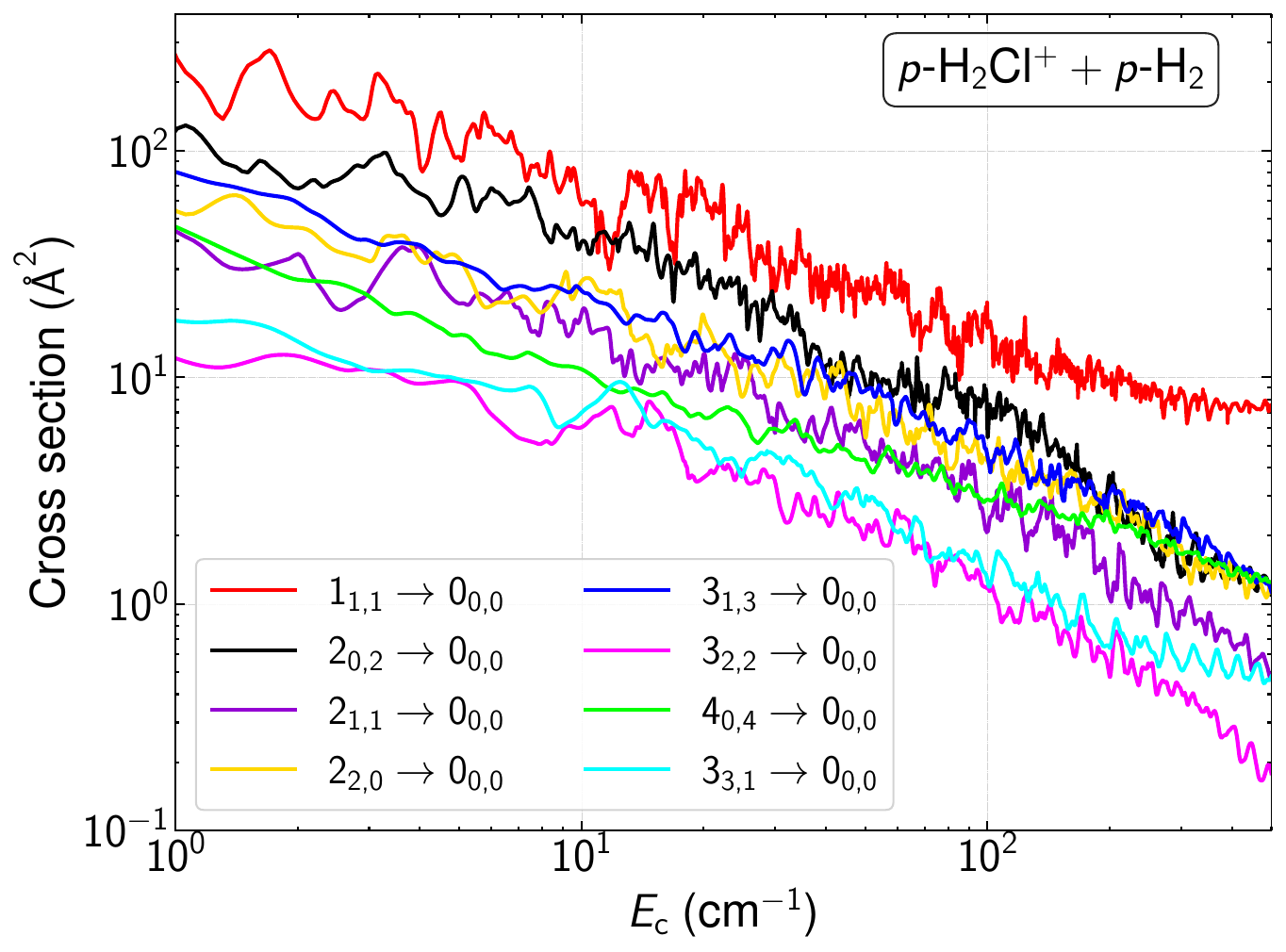} }}%
\caption{Collision-energy-dependence of the cross sections for all de-excitation transitions towards the ground rotational states of {\it ortho-}\ch{H2Cl+} (a) and {\it para-}\ch{H2Cl+} (b) due to collision with ground-state {\it para-}\ch{H2}.}
\label{fig:CS_prop}
\end{figure}

Fig.~\ref{fig:CS_prop} presents the variation of the cross sections for all transitions towards the lowest rotational state of the corresponding nuclear spin species of chloronium, \ie~towards the $1_{0,1}$ final state in the case of the {\it o}-\ch{H2Cl+} and $0_{0,0}$ in the case of the {\it p}-\ch{H2Cl+}. As can be seen, there are no strong propensity rules with respect to the $\{j,k_a,k_c\}$ quantum numbers. However, the strongest, most dominant transitions are observed in the case of $\Delta j = 0$ and $\Delta j = 1$  transitions, in accordance with the general dipole-selection rules. The same trend has been found in the case of collision with He \cite{mehnen2024}, as well as in the case of other similar asymmetric top species such as \ch{H2O} \cite{valiron2008,Faure_2007} or \ch{NH2}\cite{Bouhafs_2017} (it is worth to note that in the case of \ch{H2S} only $\Delta j = 0$ propensities were clearly found \cite{dagdigian2020}). Other propensities are found also in the case of those transitions where either $k_a$ or $k_c$ projection quantum numbers are conserved or decrease by a single value $(\Delta k_a, \Delta k_c = -1)$. It is worth mentioning that these trends are very different from those observed for a large asymmetric top species in our recent work \cite{Demes_2024}, where strong propensities have been found only in the case of $k_c$-conserving $(\Delta k_c=0)$ transitions. We examined the propensity trends for other transitions between $j=3$ and $j=2$ states, and we found that they are valid in these cases as well. It means that $\Delta j = 2$ or $\Delta k_a, k_c = 2$ or higher transitions are not favoured. Those de-excitation processes, where $k_a$ or $k_c$ increases, are not favourable neither. It can be seen from the figures that the propensity rules are more governed by the combination of the quantum numbers rather than by the internal energy difference between the initial and final states (see Table~\ref{tab:rot-levels} for details on rotational levels). More significant differences between the cross sections' magnitude are typical at lower collision energies, \ie~up to $\sim 200$~\cmmo, while most of them tend to be closer in the high-energy region, apart from the most dominant ones. In general, the scaling trends between different excitation and de-excitation channels can be better understood by finding the propensity rules.

\begin{figure}
\centering
    \subfloat[\label{subfig_a:RA_dj1o}]{{\includegraphics[width=0.49\textwidth]{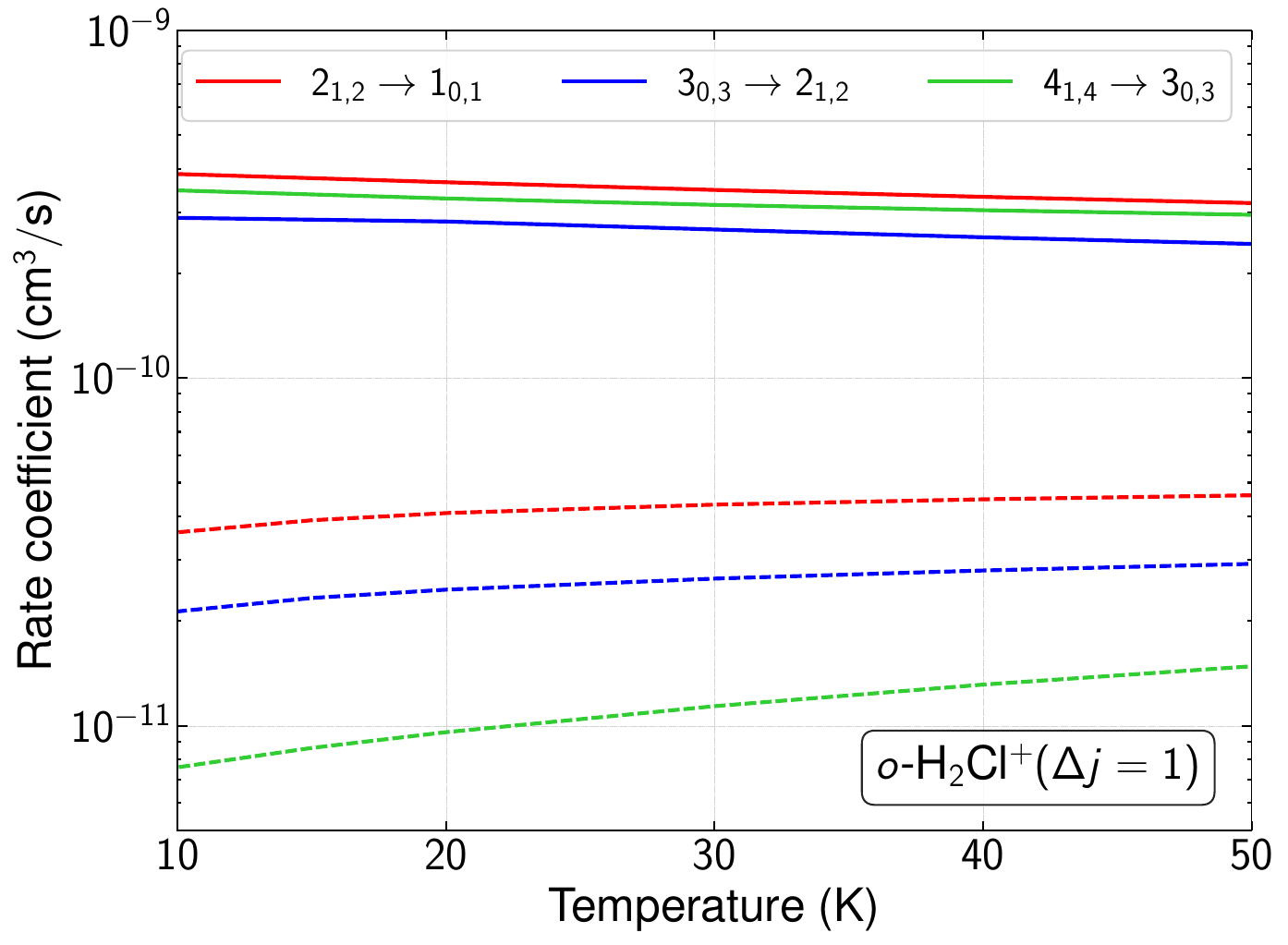} }}%
    \subfloat[\label{subfig_b:RA_dj1p}]{{\includegraphics[width=0.49\textwidth]{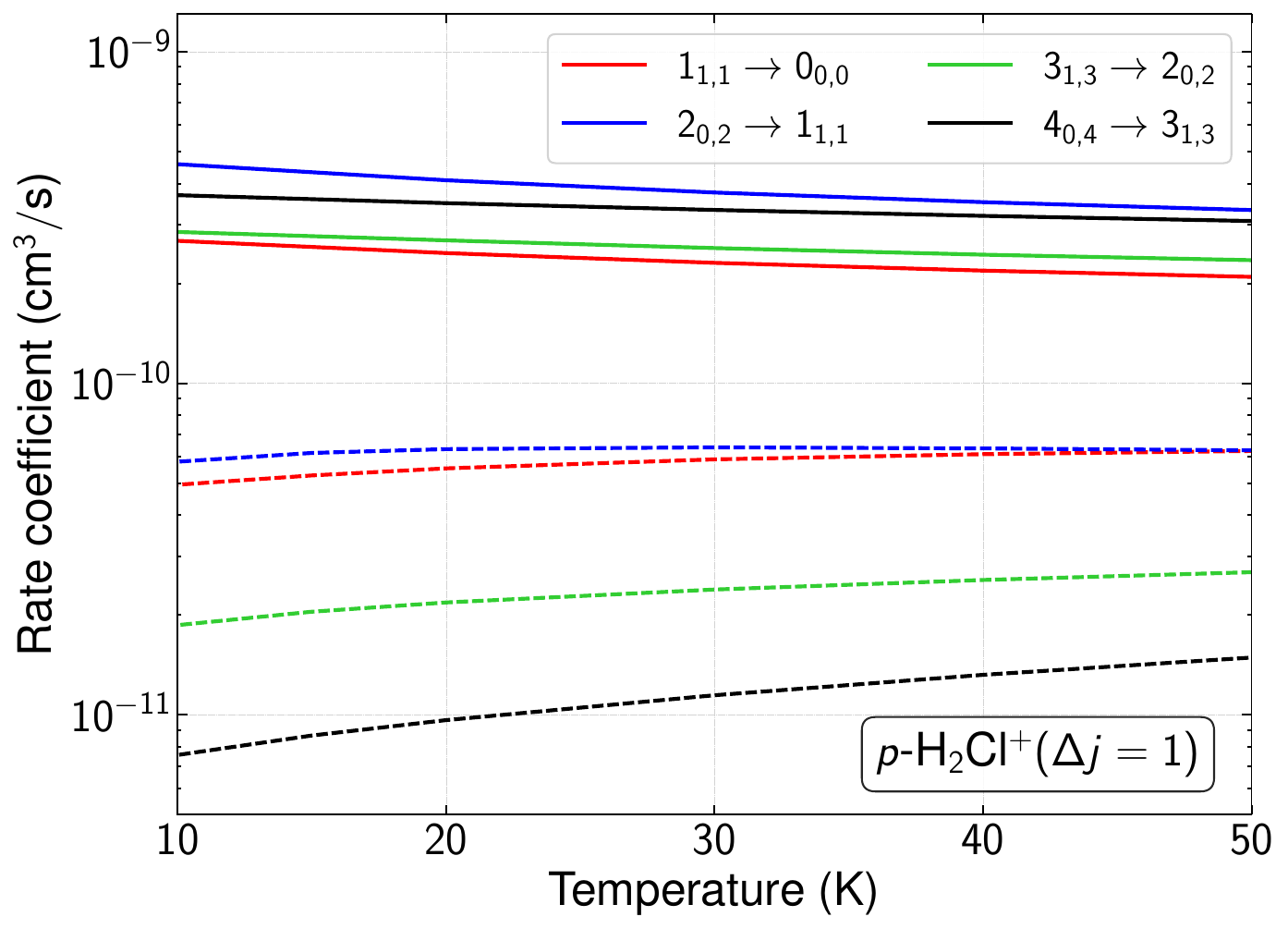} }}%
\caption{Temperature-dependence of the state-to-state rate coefficients for $\Delta j = 1$ rotational de-excitation transitions of {\it ortho-}\ch{H2Cl+} (a) and {\it para-}\ch{H2Cl+} (b) due to collision with ground-state {\it para-}\ch{H2} (solid lines, current results) and He projectiles (dash-dot lines, reproduced from \citet{mehnen2024}).}
\label{fig:RA_dj1}
\end{figure}

For a broader analysis as well as to address the possible implications of the current results in astrophysical modelling, we also examined the temperature-dependence of the state-to-state rate coefficients for the rotational de-excitation of \ch{H2Cl+}. Some $\Delta j = 1$ transitions are shown in Fig.~\ref{fig:RA_dj1}.(a)-(b) for collisions involving the {\it ortho} and {\it para} nuclear spin species of chloronium, respectively. We found some differences between the collisional data calculated with \ch{H2} and He, which can be very important in astrophysical applications. First, as expected from the magnitude of the cross sections, there are multiple factors of differences between the data that corresponds to the same transition in \ch{H2Cl+}. While in the case of the most dominant transition ($2_{1,2} \rightarrow 1_{0,1}$ or $2_{0,2} \rightarrow 1_{1,1}$) the differences are usually within factors of 6 to 8, in the case of some other processes more than an order of magnitude of deviations can be found. Extremely large gap (factors of $\sim 25-30$) is observed between the rate coefficients corresponding the $4_{1,4} \rightarrow 3_{0,3}$ and $4_{0,4} \rightarrow 3_{1,3}$ transitions. According to the calculations of \citet{mehnen2024}, these latter processes are rather negligible in collisional de-excitation due to He, while our calculations show that these are among the most significant transitions in collision with \ch{H2}. What is also obvious from the analysis is that there is no linear scaling trends between the rate coefficients obtained with the different colliding partners: while our new collisional data with \ch{H2} are systematically larger than those of with He \cite{mehnen2024}, the relative differences between them varies from transition to transition. For example, according to the authors of Ref.~\cite{mehnen2024}, the fundamental $1_{1,1} \rightarrow 0_{0,0}$ transition for {\it p-}\ch{H2Cl+} is almost as intense as the $2_{0,2} \rightarrow 1_{1,1}$, while our results show that the former process is $\sim 40\%$ weaker, most probably due to the large-amplitude resonances. Finally, it is worth to highlight another significant difference: while the collisional rate coefficients with He exhibit a systematic positive temperature gradient, the newly calculated rate coefficients show opposite trends, slight decrease with increasing temperature. These effects could have high importance in radiative transfer astrophysical modelling, especially in the case of interpreting observations of warm environments, where high-temperature rate coefficients are also involved. In a forthcoming paper, we plan to extend the range of rotational energy levels and collision energies to derive such rate coefficients for the \ch{H2Cl+ - H2} collision up to at least $200$~K kinetic temperatures, and demonstrate their impact on the astrophysical simulations.

\section{\label{sec:Concl} Conclusions}

We present an accurate 5-dimensional rigid rotor potential energy surface for studying the collisional dynamics of \ch{H2Cl+} with \ch{H2}. For the {\it ab initio} calculations, the explicitly correlated coupled cluster theory (CCSD(T)-F12b) has been used with the aug-cc-pVTZ basis set. An analytical fit has been employed then to describe the interaction potential by a set of 228 angular functions at all intermolecular distances, including anisotropies up to $l_1 = 15$ and $l_2 = 6$. The RMS and mean errors of the fit are below 1~\cmmo~both in the well and in the long-range. The calculations show a rather deep well (about $-1718$~\cmmo) along the potential energy surface, and the corresponding geometry is typical for a van der Waals complex.

The 5D PES has been implemented in the \texttt{MOLSCAT} scattering software and state-to-state cross sections have been calculated for the rotational (de-)excitation of \ch{H2Cl+} in collision with {\it p}-\ch{H2} $(j_2=0)$, using the numerically exact close-coupling scattering theory. All transitions between the lowest 9 rotational levels of both {\it ortho}- and {\it para} nuclear spin species of chloronium have been calculated (with internal energies $<125$ \cmmo), involving states with $j \leq 4$. Collision energies from $0.1$ to $500$ \cmmo~have been studied, where a dense resonance structure characterizes the cross sections for all particular channels. From the cross sections, the corresponding thermal rate coefficients have been derived up to $50$~K.

The propensity rules have also been discussed by analysing the magnitude of the cross sections for various de-excitation processes. While there is no significant dependence on the energy difference (\ie~transition frequency), the results show that the most dominant transitions are those where the changes in rotational quantum numbers obey the dipole-selection rules, in particular the ones with $\Delta j = 0, 1$ and $\Delta k = 0,1$. These propensities are equally valid both for the {\it ortho} and {\it para} nuclear spin symmetries.

Since collisional data for the \ch{H2Cl+ + H2} interaction is studied for the first time, we compared our results with the corresponding data for collision with He \citep{mehnen2024}, which is often used as a proxy for \ch{H2} in astrophysical modelling. Significant differences are found in the cross sections and rate coefficients of chloronium, which often exceeds the order of magnitude. While the collisional data are systematically larger in the case of interaction with \ch{H2}, there are no general scaling trends between the two datasets obtained with the different projectiles. It can be seen that the new collisional data should have a significant impact on astrophysical radiative transfer models and may help to better estimate the abundance of chloronium in molecular clouds, which has great implications for interstellar Cl-chemistry. Nevertheless, for an extended set of rate coefficients suitable for astrophysical applications, the energy range and the number of rotational levels should be extended, which will be addressed in a forthcoming study.

\begin{acknowledgement}

We acknowledge financial support from the European Research Council (Consolidator Grant COLLEXISM, Grant Agreement No. 811363) and the Programme National ‘Physique et Chimie du Milieu Interstellaire’ (PCMI) of CNRS/INSU with INC/INP cofunded by CEA and CNES. We also acknowledge the support from the CEA/GENCI for awarding access to the TGCC/IRENE supercomputer within the A0110413001 project. This article is based upon collaborations supported by the COST Action CA21101--Confined Molecular Systems: From a New Generation of Materials to the Stars (COSY), supported by COST (European Cooperation in Science and Technology). The authors are very grateful to P. J. Dagdigian and J. K\l{}os for the fruitful discussions and B. Mehnen for providing us the full set of collisional data for comparison. F.~L. acknowledges the Institut Universitaire de France.

\end{acknowledgement}

\begin{suppinfo}

The {\it ab initio} dataset of the \ch{H2Cl+ + H2} potential energy surface along with the expansion coefficients are provided as online supplementary material at the publisher (\url{https://doi.org/10.1021/acs.jpca.4c07467}). The full set of state-to-state rotational (de-)excitation cross sections and thermal rate coefficients are also available in the Supporting Information.

\end{suppinfo}

\bibliography{references}

\end{document}